\documentclass{ieeetj}
\usepackage{cite}
\usepackage{amsmath,amssymb,amsfonts}
\usepackage{algorithmic}
\usepackage{graphicx,color}
\usepackage{textcomp}
\usepackage{xcolor}
\usepackage{hyperref}
\hypersetup{hidelinks=true}
\usepackage{algorithm,algorithmic}
\usepackage{multirow}
\usepackage{subcaption}
\def\BibTeX{{\rm B\kern-.05em{\sc i\kern-.025em b}\kern-.08em
    T\kern-.1667em\lower.7ex\hbox{E}\kern-.125emX}}
\AtBeginDocument{\definecolor{tmlcncolor}{cmyk}{0.93,0.59,0.15,0.02}\definecolor{NavyBlue}{RGB}{0,86,125}}

\def\OJlogo{\vspace{-4pt}$<$Society logo(s) and publication title will appear here.$>$}
\def\seclogo{\vspace{10pt}$<$Society logo(s) and publication title will appear here.$>$}

\def\authorrefmark#1{\ensuremath{^{\textbf{#1}}}}

\begin{document}
\receiveddate{XX Month, XXXX}
\reviseddate{XX Month, XXXX}
\accepteddate{XX Month, XXXX}
\publisheddate{XX Month, XXXX}
\currentdate{XX Month, XXXX}
\doiinfo{XXXX.2022.1234567}

\markboth{}{Author {et al.}}

\title{Leveraging Beam Search Information for Confidence Estimation in E2E ASR}

\author{Yichen Jia\authorrefmark{1}, Hugo Van hamme\authorrefmark{1}, Senior Member, IEEE}
\affil{Department Electrical Engineering ESAT-PSI, KU Leuven, Belgium}
\corresp{Corresponding author: Yichen Jia (email: yichen.jia@kuleuven.be).}
\authornote{This research was supported by the Flemish Government under the FWO-SBO grant S004923N: NELF, and KU Leuven grant C24M/22/025.}


\begin{abstract}
To estimate confidence for end-to-end Automatic Speech Recognition (ASR) systems, recent research has proposed Confidence Estimation Modules that incorporate features from the backbone ASR model. Most existing approaches, however, are architecture-dependent. In this paper, we propose the Score-Rank Confidence Estimation Module (SR-CEM), a lightweight module that leverages beam search information to generate token- and word-level confidence scores. Specifically, SR-CEM constructs features by combining the scores and ranks of tokens within a hypothesis. Experiments show that SR-CEM achieves effective calibration on both in-domain and out-of-domain English data. On the in-domain test set, it attains a Maximum Calibration Error of 4.50\% and an Expected Calibration Error of 0.30\% at the token level, significantly outperforming softmax confidence (20.04\% and 1.75\%, respectively). At the word level, SR-CEM achieves 8.17\% and 0.35\%, compared to 17.91\% and 1.67\% from softmax confidence. Furthermore, we demonstrate its robustness across hybrid and transducer ASR architectures with different decoding strategies, as well as on Dutch, noisy and conversational speech conditions. Our main finding is that SR-CEM is particularly effective in reducing Maximum Calibration Error, which is critical for reliable downstream use of ASR outputs, while maintaining architecture independence and generality across diverse evaluation conditions.
\end{abstract}

\begin{IEEEkeywords}
Automatic Speech Recognition, Beam Search, Confidence Estimation
\end{IEEEkeywords}

\maketitle

\section{INTRODUCTION}

\IEEEPARstart{C}{ontemporary} End-to-End (E2E) Automatic Speech Recognition (ASR) systems~\cite{chan2016listen, gulati2020conformer} achieve strong recognition performance but yield biased confidence scores, typically overestimating certainty~\cite{hendrycks2016baseline, guo2017calibration}. Confidence derived from softmax probabilities~\cite{hendrycks2016baseline} remains unreliable due to model overconfidence and biases from language models, beam search, and pruning~\cite{li2021confidence}. Reliable confidence is crucial for downstream tasks such as active learning~\cite{drugman2019active}, semi-supervised learning~\cite{kreyssig2024active}, system combination~\cite{evermann2000posterior}, error correction~\cite{naderi2024towards}, and user feedback.

Prior work on confidence estimation has progressed along two main directions. \textit{Calibration methods} such as Temperature Scaling~\cite{guo2017calibration} and Matrix Scaling~\cite{kull2019beyond} apply post-hoc adjustments to model outputs using a small number of parameters. While computationally efficient, these methods assume a single calibration curve applies uniformly, limiting effectiveness when models exhibit both over- and underconfidence.

\textit{Confidence Estimation Modules (CEMs)} learn to predict confidence from model features. Li et al.~\cite{li2021confidence} proposed an MLP-based CEM for Listen, Attend and Spell using decoder hidden states and attention. Qiu et al.~\cite{qiu2021learning} extended this to word-level estimation with aggregated features including hidden states, attention distributions, and top-K scores. Wang et al.~\cite{wang2021word} developed a CEM for RNN-T incorporating acoustic and lexical features from encoder and prediction networks. Naowarat et al.~\cite{naowarat2023word} proposed letter-level features from CTC outputs. More recently, Aggarwal et al.~\cite{aggarwal2025adopting} adapted confidence estimation for Whisper, and Ravi et al.~\cite{ravi2024teles} introduced lexical similarity-based features.

Despite their effectiveness, these CEMs share a critical limitation: \textit{architecture dependence}. They rely on internal model representations (hidden states, attention weights, encoder outputs) that vary across architectures, requiring model-specific adaptations and hindering portability. Methods designed for attention-based models cannot be applied to CTC-only or transducer systems without substantial modification.

ASR decoding inherently encodes rich confidence cues in token scores and ranking. Motivated by this, we introduce the \textit{Score-Rank Confidence Estimation Module (SR-CEM)}, a lightweight model that relies solely on beam search outputs, avoiding dependence on architecture-specific representations. While prior work has used ranking~\cite{qiu2021learning} and top-K scores~\cite{qiu2021learning} as input features alongside architecture-specific representations, SR-CEM demonstrates that beam search information alone suffices for effective confidence estimation. This enables true architecture-agnostic deployment across various ASR architectures, such as hybrid CTC/Attention, attention-only, CTC-only, and RNN-T models.

Our contributions are threefold: 
1) We identify and analyze the local-global mismatch in beam search that causes systematic confidence bias, particularly for non-rank-one tokens;  
2) We propose SR-CEM, a lightweight architecture-agnostic CEM that relies solely on beam search information;  
3) We provide comprehensive evaluation demonstrating consistent improvements across architectures, languages (English, Dutch), and conditions (clean, noisy, conversational), with particular effectiveness in reducing maximum calibration error.

\section{BACKGROUND}
\subsection{Confidence in ASR}

\label{sec:confidence_asr}
In large-vocabulary subword-based ASR systems~\cite{prabhavalkar2023end}, an input sequence $X = \langle x_1, \ldots, x_M \rangle$ is mapped to a subword hypothesis $Y = \langle y_1, \ldots, y_T \rangle$, which is concatenated into a word-level transcription $W = \langle w_1, \ldots, w_L \rangle$. We use “subword” and “token” interchangeably\cite{prabhavalkar2023end}, with tokens drawn from a vocabulary $\mathcal{V}$ of size $V$.

At each decoding step $t$, the model produces a score vector $\mathbf{s}_t \in \mathbb{R}^V$. The softmax confidence for token $y_t$, $c_Y$ is
\vspace{-1mm}
\begin{equation}
    c_Y^{\text{softmax}}(y_t) = \frac{\exp(\mathbf{s}_t[y_t])}{\sum_{v=1}^V \exp(\mathbf{s}_t[v])},
\end{equation}
which we refer to as \textit{softmax} confidence. It reflects the model’s raw probability for the selected token, without calibration.

Word-level confidence scores are derived by aggregating the confidence values of the corresponding subword tokens. Specifically, for a word $w_l$ with tokens $\{y_{l,1}, \ldots, y_{l,Q_l}\}$, the word-level confidence, $c_W$ is computed as
\vspace{-1mm}
\begin{equation}
c_W(w_l) = \operatorname{agg}(c_Y(y_{l,1}), \ldots, c_Y(y_{l,Q_l})),
\end{equation}
where $\operatorname{agg}$ denotes an aggregation function, such as the mean, minimum, maximum, or product. Following the approach in~\cite{qiu2021learning}, we adopt the mean of token confidence as the confidence of a word. The mean normalizes for word length while representing the expected proportion of correct tokens.

\vspace{-2mm}
\subsection{Confidence Calibration}
\label{sec:calibration}
Confidence calibration aims to produce scores \(c \in [0,1]\) that accurately reflect prediction correctness. Let $\mathcal{D}=\{y_1,\ldots,y_I\}$ denote a set of predictions (either tokens or words), where each prediction $y_i$ is associated with a confidence $c(y_i)$ and correctness $z(y_i)\in\{0,1\}$, with $1$ denotes correct and $0$ incorrect. 

The predictions are partitioned into bins based on their confidence scores using adaptive binning~\cite{ding2020revisiting}. Unlike fixed binning schemes that pre-specify the number of bins \(K\) and their boundaries, adaptive binning dynamically determines both the number of bins and their boundaries based on the data distribution.
Let $B_k$ denote the $k$-th bin for $k=1,\ldots,K$, where $K$ is data-dependent, and let $D_k = \{y_i \mid c(y_i)\in B_k\}$ represent the set of predictions assigned to this bin.

We define the joint distribution over confidence and correctness as $P_{\theta,\mathcal{D}}(c,z)$, where $\theta$ are the model parameters.  
Calibration error is quantified by Expected Calibration Error (ECE) and Maximum Calibration Error (MCE):
\vspace{-1mm}
\begin{equation}
    \widehat{\mathrm{ECE}} = \frac{1}{|\mathcal{D}|} \sum_{k=1}^K \left| \sum_{y_i \in D_k} c(y_i) - \sum_{y_i \in D_k} z(y_i) \right|
    \label{eq:ECE}
\end{equation}
\vspace{-1mm}
\begin{equation}
    \widehat{\mathrm{MCE}} = \max_k \frac{1}{|D_k|} \left| \sum_{y_i \in D_k} c(y_i) - \sum_{y_i \in D_k} z(y_i) \right|
    \label{eq:MCE}
\end{equation}

\begin{figure}[tb]
    \vspace{-3mm}
    \centering
    \begin{subfigure}[b]{0.49\linewidth}
        \includegraphics[width=\linewidth]{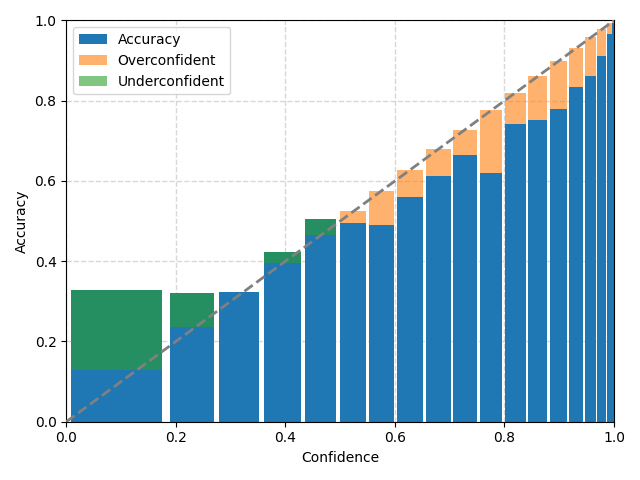}
        \caption{Token level}
        \label{fig:softmax_token}
    \end{subfigure}
    \hfill
    \begin{subfigure}[b]{0.49\linewidth}
        \includegraphics[width=\linewidth]{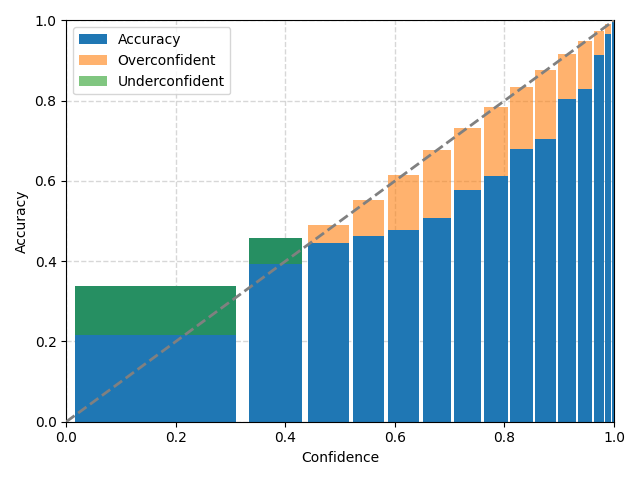}
        \caption{Word level}
        \label{fig:softmax_word}
    \end{subfigure}
    \caption{Reliability diagrams for softmax confidence scores on LibriSpeech test-clean.}
    \label{fig:softmax}
    \vspace{-2mm}
\end{figure}

Reliability diagrams~\cite{guo2017calibration} visualize calibration by comparing confidence and accuracy per bin. Fig.~\ref{fig:softmax} shows token- and word-level diagrams for softmax confidence on LibriSpeech test-clean~\cite{panayotov2015librispeech}. Overconfidence is a well-known issue in deep learning models~\cite{guo2017calibration}, shown by the orange bars, whereas the green bars reflect underconfidence. With adaptive binning, both the number of bins and their characteristics (width and population) are determined by the algorithm. Wider bins may indicate regions where calibration is more uniform, while narrower bins capture regions requiring finer granularity for accurate calibration assessment.

\vspace{-2mm}
\subsection{Beam Search and Scoring in ASR}
\label{sec:beam_serach_scoring}

Many modern ASR architectures use beam search~\cite{watanabe2017hybrid}, where hypotheses are expanded token by token and ranked by cumulative log probability scores. Beam search uses scores to select tokens, making them a natural representation of confidence. However, a critical mismatch exists: softmax confidence reflects local probability $p(y_t | y_{<t}, X)$, while beam search selects based on cumulative score. This causes systematic underconfidence for tokens with low local probability that appear in the best hypothesis.

Let \(r(y_t)\) denote the rank of \(y_t\) in score vector \(\mathbf{s}_t\) (\(r=1\) is highest). Softmax confidence satisfies:
\vspace{-1mm}
\begin{equation}
   c_Y^{\text{softmax}}(y_t) = \frac{\exp(\mathbf{s}_t[y_t])}{\sum_{v=1}^V \exp(\mathbf{s}_t[v])} \leq \frac{1}{r(y_t)},
   \label{eq:rank_bound}
\end{equation}
with equality only if top-\(r\) scores are identical. In practice, \(c_Y^{\text{softmax}}(y_t)\) is well below this bound, making correct non-rank-one tokens systematically underconfident (Fig.~\ref{fig:softmax}). This motivates our approach: SR-CEM uses rank to capture local competition, context scores for global quality, and top-K scores for uncertainty among alternatives. These features depend only on beam search dynamics, not internal model representations, enabling cross-architecture generalization while potentially being less effective with aggressive pruning or constrained decoding that limits alternative exploration.


\section{METHODOLOGY}


Following the CEM framework, we construct features from beam search (token ranks and scores) and feed them into a lightweight module to predict confidence \(\hat{c}\). We call this the \textit{Score-Rank Confidence Estimation Module (SR-CEM)}.

\vspace{-2mm}
\subsection{Token-Level SR-CEM}
We design SR-CEM as a lightweight module consisting of one hidden layer with Rectified linear unit (ReLU) activation and a sigmoid output. The predicted confidence for token \(y_t\) is
\vspace{-1mm}
\begin{equation}
    \hat{c}_Y(y_t) = \sigma\!\left(W_{Y,2} \cdot \text{ReLU}(W_{Y,1} \mathbf{f}_Y(y_t) + \mathbf{b}_{Y,1}) + \mathbf{b}_{Y,2}\right),
    \label{eq:SR_CEM_token}
\end{equation}
\vspace{-1mm}
where \(\sigma(\cdot)\) is the sigmoid, \(W_{Y,1} \in \mathbb{R}^{64 \times d_Y}\), \(W_{Y,2} \in \mathbb{R}^{1 \times 64}\), and \(d_Y\) is the input feature dimension. The token-level feature vector $\mathbf{f}_Y(y_t)$ is described in the following text.

At each decoding step \(t\), let \(\mathbf{s}_t \in \mathbb{R}^V\) denote the score vector over the vocabulary \(\mathcal{V}\) of size \(V\). The selected token is \(y_t \in \mathcal{V}\) with score \(s_Y(y_t) = \mathbf{s}_t[y_t]\). The final hypothesis is \(Y = \langle y_1, \ldots, y_T \rangle\), with total score
\vspace{-1mm}
\begin{equation}
    S(Y) = \sum_{t=1}^T s_Y(y_t).
\end{equation}
For each token, the score $\mathbf{s}_t[y_t]$ provides basic confidence information. However, as discussed earlier, this score alone may be insufficient, since other tokens in the hypothesis also carry relevant information. To incorporate context, we define cumulative scores of preceding and succeeding tokens:
\vspace{-1mm}
\begin{equation}
    S_{<t} = \sum_{k=1}^{t-1} s_Y(y_k), \qquad 
    S_{>t} = \sum_{k=t+1}^{T} s_Y(y_k),
\end{equation}
so that \(S(Y) = S_{<t} + s_Y(y_t) + S_{>t}\).
Let \(r(y_t)\) be the rank of \(y_t\) within \(\mathbf{s}_t\). Rank provides additional information beyond the raw softmax confidence. Following~\cite{qiu2021learning}, we also include \(\text{Topk}(t) \in \mathbb{R}^K\), the top-\(K\) scores in \(\mathbf{s}_t\) with \(K=4\), which capture uncertainty and competition among alternatives.
We therefore define the token-level feature vector as
\vspace{-1mm}
\begin{equation}
    \mathbf{f}_Y(y_t) = \big[s_Y(y_t),\ r(y_t),\ S_{<t},\ S_{>t},\ \text{Topk}(t)\big],
\end{equation}
where the 8 dimensions comprise: token score (1), token rank (1), preceding cumulative scores (1), succeeding cumulative scores (1), and top-k scores (4).


\vspace{-2mm}
\subsection{Word-Level SR-CEM}

We design SR-CEM to predict word-level confidence using a single hidden layer with ReLU activation followed by a sigmoid output:
\vspace{-1mm}
\begin{equation}
    \hat{c}_W(w_l) = \sigma\!\left(W_{W,2} \cdot \text{ReLU}(W_{W,1} \mathbf{f}_W(w_l) + \mathbf{b}_{W,1}) + \mathbf{b}_{W,2}\right),
    \label{eq:SR_CEM_word}
\end{equation}
where \(\sigma(\cdot)\) is the sigmoid, \(W_{W,1} \in \mathbb{R}^{64 \times d_W}\), \(W_{W,2} \in \mathbb{R}^{1 \times 64}\), and \(d_W\) is the input feature dimension. The word-level feature vector $\mathbf{f}_W(w_l)$ is described in the following text.

To directly predict word-level confidence, we construct features from the corresponding token-level features. Let \(w_l\) denote the \(l\)-th word in the hypothesis, consisting of \(Q_l\) tokens:  
\( w_l = \langle y_{l,1}, \ldots, y_{l,Q_l} \rangle \),  
where each token \(y_{l,j} \in \mathcal{V}\) has an associated score vector \(\mathbf{s}_{l,j} \in \mathbb{R}^V\).

The word score is defined as the sum of the token scores:
\vspace{-1mm}
\begin{equation}
    s_W(w_l) = \sum_{j=1}^{Q_l} \mathbf{s}_{l,j}[y_{l,j}].
\end{equation}

Analogous to the token-level case, we define preceding and succeeding scores:
\vspace{-1mm}
\begin{equation}
    S_{<l} = \sum_{k=1}^{l-1} s_W(w_k), 
    \qquad
    S_{>l} = \sum_{k=l+1}^{L} s_W(w_k),
\end{equation}
where \(L\) is the total number of words in the hypothesis.

The word rank \(r_l\) is taken as the maximum token rank within the word:
\vspace{-1mm}
\begin{equation}
    r_l(w_l) = \max_{1 \leq j \leq Q_l} r(y_{l,j}),
\end{equation}
where \(r(y_{l,j})\) is the rank of token \(y_{l,j}\) in \(\mathbf{s}_{l,j}\). Among several aggregation strategies (mean, min, sum), the maximum rank performed best in preliminary experiments.

We define the word-level feature vector as
\vspace{-1mm}
\begin{equation}
    \mathbf{f}_W(w_l) = \big[s_W(w_l),\, r_l(w_l),\, S_{<l},\, S_{>l},\, Q_l\big],
\end{equation}
where the 5 dimensions comprise: word score (1), word rank (1), preceding cumulative scores (1), succeeding cumulative scores (1), and token count (1).

\vspace{-2mm}
\subsection{Data Preparation and Training}

Informative features are collected during decoding for each token in the training data. The key input is the token score vector $\mathbf{s}_t$ at each decoding step of the winning hypothesis. After decoding, token and word correctness are labeled using Levenshtein alignment between ground-truth and predicted transcriptions.

Recognition errors include insertion, deletion, and substitution. Since deletions correspond to missing units, no confidence score can be assigned, and they are excluded from experiments. Remaining units are labeled as correct ($z=1$) or incorrect ($z=0$), with insertions and substitutions marked incorrect.

Training data are thus pairs of feature vectors and correctness labels. The objective is Binary Cross-Entropy (BCE) loss between the predicted confidence $\hat{c}$ and label $z$:
\vspace{-1mm}
\begin{equation}
    \mathcal{L}_{\text{BCE}}(\hat{c}, z) = - \left[ z \cdot \log(\hat{c}) + (1 - z) \cdot \log(1 - \hat{c}) \right].
    \label{eq:BCE}
\end{equation}

\section{EXPERIMENTAL SETUP}

Experiments are conducted with ESPnet~\cite{watanabe2018espnet}, using custom code to extract decoding-time features without altering the original workflow. The proposed SR-CEM and baselines are implemented in PyTorch~\cite{paszke2019pytorch}. Code to reproduce the experiments will be made available.\footnote{\url{https://github.com/windskylionheart1023/Score_Rank_Confidence_Estimation_Module}}


\vspace{-2mm}
\subsection{Datasets and Backbone ASR}
\label{sec:backbone_asr}



\textit{LibriSpeech}~\cite{panayotov2015librispeech} (LS) is a 1,000-hour English audiobook corpus with \textit{clean} and \textit{other} test splits. \textit{Common Voice}~\cite{ardila2019common} (CV) is a multilingual, crowd-sourced corpus with diverse accents. Due to transcription variability, ASR performance on CV is lower than LS~\cite{radford2023robust,peng2023reproducing}, making confidence estimation more challenging. We use only the U.S. accent subset from CV.

\textit{Libri-Adapt}~\cite{mathur2020libri} (APT) is a benchmark dataset for domain adaptation in ASR. It extends LibriSpeech by introducing two dimensions of variability: microphone type and noise condition, while keeping the original transcriptions unchanged. The dataset includes six types of microphones and four types of background noise.

\textit{Corpus Gesproken Nederlands}~\cite{oostdijk2000spoken} (CGN) is a Dutch speech corpus covering spontaneous speech, interviews, and read speech from both the Netherlands and Flanders. We use the Flemish part with the train, test, and validation splits from~\cite{poncelet2025leveraging} to evaluate cross-lingual robustness of confidence estimation.

\textit{CHiME-6}~\cite{watanabe2020chime} is a conversational speech corpus featuring spontaneous dinner party conversations with background noise. It provides a more realistic and challenging benchmark than read or prompted speech datasets.

In our experiments, we treat LS as the main dataset and use CV as a domain-mismatch dataset, as both the audio characteristics and transcripts differ substantially. When an ASR model is trained on LS and evaluated on CV, the performance degradation is even more pronounced. The APT corpus is used as the noisy dataset, where we fine-tune the ASR model and subsequently train and evaluate the CEMs. We use CHiME-6 as the conversational speech dataset, where we also fine-tune the ASR model on it, then we train and evaluate the CEMs.

The backbone ASR system is a \textit{hybrid CTC/Attention} model~\cite{watanabe2017hybrid} with 12 Conformer encoder and 6 Transformer decoder layers. Both use an attention dimension of 256, feed-forward dimension of 2048, and 4 attention heads. The CTC weight is 0.3.
The model is trained from scratch on LS train-clean-360 for 40 epochs with Adam~\cite{kingma2015adam}, using a learning rate of 4.0, achieving 4.6\% WER and 5.6\% TER on LS test-clean.
For out-of-domain evaluation, we test on CV (U.S. English subset: 45.8\% WER, 48.3\% TER).

We employ an \textit{RNN-Transducer (RNN-T)} model~\cite{graves2012sequence}, in which the acoustic and prediction networks jointly generate scores for each encoder frame.
It is a Conformer-Transducer model with 12 Conformer encoder blocks and a single-layer LSTM prediction network. The encoder uses an attention dimension of 256, feed-forward dimension of 2048, and 4 attention heads, with relative positional encoding, and convolution modules enabled. The joint network projects into a 320-dimensional space. A hybrid training objective is used with a CTC weight of 0.3.
The RNN-T model is also trained from scratch on LS train-clean-360 for 40 epochs using Adam~\cite{kingma2015adam} (learning rate=4.0), with 6.1\% WER and 7.5\% TER on LS test-clean.

The vocabulary $\mathcal{V}$ (Sec.~\ref{sec:confidence_asr}) is built with Byte Pair Encoding on LS train-clean-360, with size 5{,}000.

\vspace{-2mm}
\subsection{Baselines}



Most existing methods are architecture-specific. Methods like~\cite{naowarat2023word} construct features with vocabulary-dependent dimensionality (impractical for our 5,000-unit vocabulary) and target CTC-only systems (our hybrid model uses CTC weight 0.3). Temperature Scaling~\cite{guo2017calibration} applies fixed adjustments unsuitable for systems with both over- and underconfidence.

We adopt: \textit{Softmax}, \textit{MLPCEM} and \textit{Xformer} (adapted from~\cite{qiu2021learning}), \textit{E2EXformer}~\cite{qiu2021learning}, and \textit{TruCLeS}~\cite{ravi2024teles} for CTC/RNN-T models.


For MLPCEM we follow~\cite{qiu2021learning}, using three fully connected layers of sizes 128, 32, and 1. As configurations for Xformer and E2EXformer were not released, we explored several setups and selected compact variants: Xformer uses hidden dim 64, feed-forward dim 128, and 1 attention head; E2EXformer uses the same hidden dim with a single Transformer decoder layer. Model sizes are reported in Tab.~\ref{tab:model_size}.

All methods are trained with Adam using a learning rate of 0.001 and weight decay of $1\times10^{-4}$, batch size 128, and early stopping. Each dataset is split 8:2 into train/validation using a fixed seed. Training runs for up to 20 epochs, and the best validation model is selected.

All the aforementioned baselines rely on a decoder with hidden states and are therefore not applicable to CTC-only models. For CTC-only and RNN-T models, we adopt \textit{TruCLeS}~\cite{ravi2024teles} as the baseline. TruCLeS is a recent CEM that leverages lexical similarity scores between hypotheses and reference candidates to enhance confidence estimation.
Because the backbone ASR model has different hidden dimensions, we adopt the original setting for TruCLeS RNN-T. The LSTMs are configured with 256 hidden units per direction and an input dimension of 1024.
The TruCLeS models are trained using the same framework described in the previous paragraph. The learning rate is set to 0.0001, following the original work. The number of parameters is 5.4M for TruCLeS-RNNT, 7.4M for TruCLeS-CTC. 

\begin{table}[tb]
\vspace{-2mm}
\caption{Number of parameters per model. For word-level MLPCEM and Xformer, sizes match token-level variants.}
\begin{tabular}{c|ccc|cc}
\hline
     & \multicolumn{3}{c|}{Token level} & \multicolumn{2}{c}{Word level} \\ \cline{2-6} 
     & MLPCEM & Xformer & SR-CEM & E2EXformer & SR-CEM \\ \hline
Size & 70k & 83k & 0.6k & 100k & 0.4k \\ \hline
\end{tabular}
\label{tab:model_size}
\vspace{-2mm}
\end{table}

\subsection{Efficiency and Deployment}

SR-CEM's two-layer architecture with 0.6k (token) / 0.4k (word) parameters is 100-250× smaller than baselines. Model memory footprint is less than 5KB. Feature storage scales with utterance length (approximately 12KB for 150 tokens), which is modest compared to ASR memory requirements. Training converges in a few minutes (20 epochs on 20k utterances, single 3060 GPU). Inference adds less than 0.1ms latency per utterance since scores and ranks are already computed during beam search.

We recommend hidden dimension $h=64$ (balances expressiveness and simplicity), learning rate 0.001, weight decay $10^{-4}$, and batch size 128. When training data contains predominantly well-calibrated high-confidence samples, consider using stratified or balanced sampling to ensure diverse batch compositions for more effective training.

\vspace{-2mm}
\subsection{Beam search for RNN-T model}
The seminal RNN-T work~\cite{graves2012sequence} describes inference as involving a summarization over different prefixes. In contrast, modern speech recognition toolkits, such as ESPnet~\cite{watanabe2018espnet}, SpeechBrain~\cite{ravanelli2021speechbrain}, NeMo~\cite{kuchaiev2019nemo}, and fairseq~\cite{ott2019fairseq}, perform beam search for RNN-T models without prefix search\cite{rao2017exploring}. Concretely, for each encoder frame, when a new non-blank token is appended to a hypothesis, a score vector is produced that assigns probabilities to all tokens in the vocabulary. If the blank token is selected, no token is added to the hypothesis. Our proposed SR-CEM is designed for this beam search setting without prefix search, which is the widely adopted approach in modern speech processing toolkits.

\vspace{-2mm}
\subsection{Evaluation Metrics}

Following~\cite{qiu2021learning, aggarwal2025adopting, naderi2024towards, naowarat2023word, li2021confidence}, we use several metrics.  
\textit{Normalized Cross Entropy (NCE)}~\cite{siu1997improved} evaluates how well scores align with correctness, ranging from $-\infty$ to 1 (higher is better).  

To assess discriminative ability (separating correct from incorrect predictions), we report \textit{AUC-ROC} and \textit{AUC-PR} (denoted ROC and PR), both between 0 and 1, where higher is better. These metrics do not measure calibration quality but are included for comparison with prior work.

Calibration is measured by \textit{Expected Calibration Error (ECE)} and \textit{Maximum Calibration Error (MCE)} (Sec.~\ref{sec:calibration}), computed using adaptive binning~\cite{ding2020revisiting} to counter the skew of high-confidence scores. While ECE reflects average miscalibration, it is often dominated by already reliable predictions. We emphasize MCE, which captures worst-case errors in rare bins, which is crucial for downstream decisions where even small subsets of poorly calibrated scores can cause significant harm.
Both ECE and MCE are reported in the tables as percentages for clarity.

\section{RESULTS AND DISCUSSION}
This section is organized as follows. We first present preliminary results with the hybrid CTC/Attention model on the in-domain and out-of-domain datasets. We then conduct an ablation study based on these experiments. Next, we extend the analysis to additional ASR architectures, including attention-only, CTC-only, and RNN-T models. We further evaluate our method on an out-of-domain dataset and on Dutch to explore language robustness. Finally, we assess the effectiveness of SR-CEM on noisy and conversational data.


\vspace{-2mm}
\subsection{Token-Level Results}

\begin{figure}[tb]
    \centering
    \begin{subfigure}[b]{0.49\linewidth}
        \includegraphics[width=\linewidth]{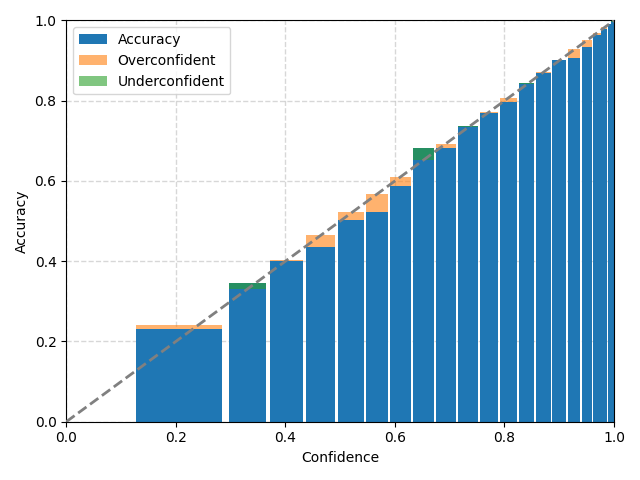}
        \caption{Token level}
        \label{fig:rd_sr_cem_token}
    \end{subfigure}
    \hfill
    \begin{subfigure}[b]{0.49\linewidth}
        \includegraphics[width=\linewidth]{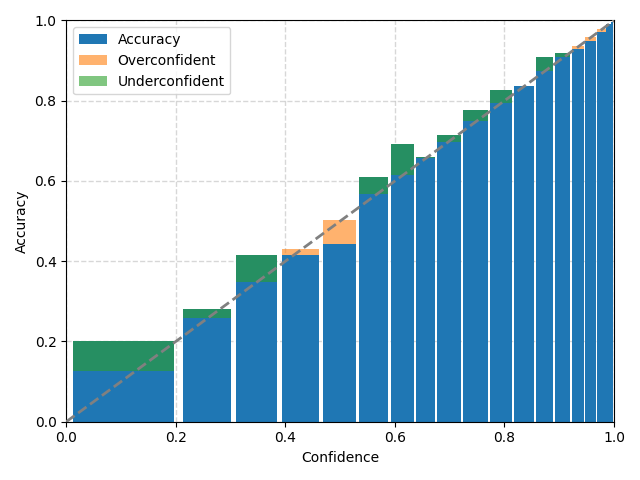}
        \caption{Word level}
        \label{fig:rd_sr_cem_word}
    \end{subfigure}
    \caption{Reliability diagrams for SR-CEM confidence estimation on LibriSpeech test-clean at token and word levels (left and right respectively). SR-CEM trained on the LibriSpeech train-clean-100.}
    \label{fig:rd_sr_cem}
    \vspace{-2mm}
\end{figure}

Tab.~\ref{tab:token_confidence} shows SR-CEM outperforms all baselines on LS test-clean, achieving MCE of 4.50\% compared to 20.04\% for softmax (see Fig.~\ref{fig:rd_sr_cem_token}). Among baselines, MLPCEM improves NCE and ECE, while Xformer performs better on MCE. On out-of-domain CV, SR-CEM maintains the best performance (MCE 19.59\%) despite poor ASR performance, while MLPCEM and Xformer achieve only 25.62\% and 29.65\%.


\begin{table}[tb]
\vspace{-2mm}
\centering
\caption{Token-level confidence results on LS test-clean and CV test sets. All CEMs trained on LS train-clean-100. ASR trained on LS train-clean-360}
\begin{tabular}{l|l|rrrrr}
\hline
Dataset     & Method     & NCE$\uparrow$ & ROC$\uparrow$ & PR$\uparrow$ & ECE$\downarrow$ & MCE$\downarrow$ \\ \hline
\multirow{4}{*}{LS} 
            & Softmax    & 0.301        & 0.919             & \textbf{0.996}            & 1.75          & 20.04         \\
            & MLPCEM        & 0.320        & 0.891             & 0.993            & 0.43          & 21.22         \\
            & Xformer    & 0.298        & 0.875             & 0.993            & 0.37          & 12.06         \\
            & \textit{SR-CEM}     & \textbf{0.383} & \textbf{0.923}   & \textbf{0.996}   & \textbf{0.30} & \textbf{4.50} \\ \hline
\multirow{4}{*}{CV} 
            & Softmax    & -0.029        & 0.853             & 0.926            & 15.99         & 34.77         \\
            & MLPCEM        & 0.126        & 0.840             & 0.911            & 12.91         & 25.62         \\
            & Xformer    & 0.125        & 0.836             & 0.913            & 13.48         & 29.65         \\
            & \textit{SR-CEM}     & \textbf{0.248} & \textbf{0.866}   & \textbf{0.931}   & \textbf{9.54}  & \textbf{19.59}\\ \hline
\end{tabular}
\label{tab:token_confidence}
\vspace{-2mm}
\end{table}

\begin{table}[tb]
\vspace{-2mm}
\centering
\caption{Word-level confidence estimation results on LS test-clean and CV
test sets. All CEMs are trained on the LS train-clean-100. ASR trained on LS train-clean-360}
\begin{tabular}{l|l|rrrrr}
\hline
Dataset     & Method       & NCE$\uparrow$ & ROC$\uparrow$ & PR$\uparrow$ & ECE$\downarrow$ & MCE$\downarrow$ \\ \hline
\multirow{5}{*}{LS} 
            & Softmax      & 0.336        & \textbf{0.931}     & \textbf{0.996}    & 1.67           & 17.91          \\
            & MLPCEM       & 0.337        & 0.899              & 0.995             & 0.55           & 16.87          \\
            & Xformer      & 0.314        & 0.886              & 0.994             & 0.71           & 20.57          \\
            & E2EXformer   & 0.325        & 0.912              & \textbf{0.996}             & 0.92           & 23.29          \\
            & \textit{SR-CEM}       & \textbf{0.356} & 0.899            & 0.994             & \textbf{0.35}  & \textbf{8.17}  \\ \hline
\multirow{5}{*}{CV} 
            & Softmax      & -0.485       & \textbf{0.792}     & \textbf{0.843}    & 23.84          & 47.21          \\
            & MLPCEM       & -0.278       & 0.774              & 0.821             & 21.07          & 41.67          \\
            & Xformer      & -0.225       & 0.783              & 0.839    & 21.94          & 43.56          \\
            & E2EXformer   & -0.327       & 0.772              & 0.830             & 20.88          & 34.87          \\
            & \textit{SR-CEM}       & \textbf{-0.198} & 0.778           & 0.804             & \textbf{18.62} & \textbf{32.78} \\ \hline
\end{tabular}
\label{tab:word_confidence}
\vspace{-2mm}
\end{table}

\vspace{-2mm}
\subsection{Word-Level Results}

Tab.~\ref{tab:word_confidence} shows SR-CEM achieves the best word-level calibration on LS test-clean, reducing MCE from 17.91\% (softmax) to 8.17\% with lowest ECE (0.35\%). While softmax achieves higher ROC/PR (0.931/0.996), SR-CEM provides better-calibrated scores (0.899/0.994) for threshold-based applications. On out-of-domain CV, SR-CEM reduces MCE from 47.33\% to 32.78\% despite all methods showing negative NCE. The word-level performance gap compared to token level (8.17\% vs. 4.50\% MCE) reflects the absence of top-K features, which limits feature expressiveness but maintains practical calibration quality.

\vspace{-2mm}
\subsection{Ablation study}

We conducted a Leave-One-Feature-Out ablation study on the LibriSpeech test-clean. As shown in Tab.~\ref{tab:ablation_token} and Tab.~\ref{tab:ablation_word}, each feature contributes to performance gains, and removing any feature results in degradation. At the token level, removing Succeeding Score Sum has the largest impact on MCE, while removing rank shows the least effect. At the word level, removing the token count has the largest impact on MCE, while removing max rank shows the least effect.

\begin{table}[tb]
\vspace{-2mm}
\centering
\caption{Leave-One-Feature-Out ablation study for token level on LS test-clean.}
\begin{tabular}{l|rrrrr}
\hline
Deleted feature         & NCE$\uparrow$ & ROC$\uparrow$ & PR$\uparrow$ & ECE$\downarrow$ & MCE$\downarrow$  \\ \hline
Score          & 0.373  & 0.919 & 0.995 & 0.53 & 5.36 \\
Rank           & 0.378  & 0.921 & \textbf{0.996} & 0.45 & 5.53 \\
Prec Score Sum & 0.377  & 0.920 & \textbf{0.996} & \textbf{0.30} & 6.36 \\
Succ Score Sum & 0.377 & 0.920 & \textbf{0.996} & 0.40 & 7.39 \\
TopK           & 0.342 & 0.898 & 0.993 & 0.60 & 5.76 \\
\textit{None}           & \textbf{0.383}  & \textbf{0.923} & \textbf{0.996} & \textbf{0.30} & \textbf{4.50} \\ \hline
\end{tabular}
\label{tab:ablation_token}
\vspace{-2mm}
\end{table}

\begin{table}[tb]
\vspace{-2mm}
\centering
\caption{Leave-One-Feature-Out ablation study for word level on LS test-clean.}
\begin{tabular}{l|rrrrr}
\hline
Deleted feature         & NCE$\uparrow$ & ROC$\uparrow$ & PR$\uparrow$ & ECE$\downarrow$ & MCE$\downarrow$  \\ \hline
Score          & 0.065  & 0.605 & 0.974 & \textbf{0.35} & 9.48 \\
Max Rank       &0.355   &0.898  &\textbf{0.994}  &0.37  &8.93  \\
Prec Score Sum &\textbf{0.356}   &0.898  &\textbf{0.994}  &0.38  &10.44  \\
Succ Score Sum &0.355  &0.898  &\textbf{0.994}  &0.45  &11.87  \\
\#Token in word              &0.349  &0.897  &\textbf{0.994}  &0.35  &17.21  \\
\textit{None}           & \textbf{0.356}  & \textbf{0.899} & \textbf{0.994} & \textbf{0.35} & \textbf{8.17} \\ \hline
\end{tabular}
\label{tab:ablation_word}
\vspace{-2mm}
\end{table}

\vspace{-2mm}
\subsection{Experiemtn with Attention-Only Decoding}

The evaluation metrics depend on both predicted confidence scores and correctness labels. In hybrid ASR, correctness is determined by the final hypothesis combining attention and CTC scorers, while all baseline methods use only attention-based scoring. For fair comparison, we conduct an additional experiment with attention-only decoding, disabling the CTC scorer, extracting corresponding features, and retraining all models.



Tab.~\ref{tab:att_confidence} shows SR-CEM outperforms baselines at token level (MCE 8.12\% vs. 16.11\% softmax) and maintains lowest word-level MCE (13.08\% vs. 21.23\% softmax), despite E2EXformer achieving competitive overall performance. While baselines benefit from the simplified setting, SR-CEM remains most effective at reducing maximum calibration error.

\begin{table}[tb]
\vspace{-2mm}
\centering
\caption{Token- and word-level confidence results on the LS test-clean with attention-only decoding. All CEMs trained on the LS train-clean-100. ASR trained on LS train-clean-360.}
\begin{tabular}{l|l|rrrrr}
\hline
Level   & Method       & NCE$\uparrow$ & ROC$\uparrow$ & PR$\uparrow$ & ECE$\downarrow$ & MCE$\downarrow$ \\ \hline
\multirow{4}{*}{Token} 
        & Softmax      & 0.264        & 0.861             & 0.989             & 1.47          & 16.11         \\
        & MLPCEM       & 0.264        & 0.856             & 0.989             & 0.78          & 23.17         \\
        & Xformer      & 0.263        & 0.845             & 0.989             & \textbf{0.47}          & 14.60         \\
        & \textit{SR-CEM}       & \textbf{0.289} & \textbf{0.864}   & \textbf{0.990}    & 0.73 & \textbf{8.12} \\ \hline
\multirow{5}{*}{Word} 
        & Softmax      & \textbf{0.351} & \textbf{0.919}   & \textbf{0.996}    & 1.14          & 21.23         \\
        & MLPCEM       & 0.329        & 0.896             & 0.995             & \textbf{0.53}         & 23.55         \\
        & Xformer      & 0.316        & 0.884             & 0.994             & 0.74          & 14.44         \\
        & E2EXformer   & 0.317        & 0.909             & \textbf{0.996}    & 0.96 & 28.93         \\
        & \textit{SR-CEM}       & 0.324        & 0.880             & 0.992             & 0.92          & \textbf{13.08}\\ \hline
\end{tabular}
\label{tab:att_confidence}
\vspace{-2mm}
\end{table}

\vspace{-2mm}
\subsection{Experiment with the CTC-Only Decoding}

In the CTC-only setup, correctness is determined directly from the CTC best path without attention scoring. Since other baselines are designed for attention-based decoding, we compare SR-CEM only with softmax confidence and the TruCLeS baseline.  



Tab.~\ref{tab:ctc_confidence} shows CTC-only results comparing SR-CEM with softmax and TruCLeS. SR-CEM achieves best token-level performance (MCE 3.27\% vs. 15.65\% softmax, NCE 0.380 vs. 0.329) and word-level performance (MCE 7.36\% vs. 18.95\% softmax, 10.6\% TruCLeS). While TruCLeS improves calibration over softmax, it shows weak discrimination, whereas SR-CEM balances both effectively.

\begin{table}[tb]
\vspace{-2mm}
\centering
\caption{Token and word-level confidence results on the LS test-clean with CTC-only decoding. All CEMs trained on the LS train-clean-100. ASR trained on LS train-clean-360.}
\begin{tabular}{l|l|rrrrr}
\hline
Level   & Method       & NCE$\uparrow$ & ROC$\uparrow$ & PR$\uparrow$ & ECE$\downarrow$ & MCE$\downarrow$ \\ \hline
\multirow{2}{*}{Token} 
        & Softmax      & 0.329 & 0.925 & \textbf{0.996} & 1.40 & 15.65 \\
        & \textit{SR-CEM} & \textbf{0.380} & \textbf{0.926} & \textbf{0.996} & \textbf{0.26} & \textbf{3.27} \\ \hline
\multirow{3}{*}{Word} 
        & Softmax      & 0.326 & 0.929 & \textbf{0.996} & 2.08 & 18.95 \\
        & TruCLeS       & 0.181 & 0.812 & 0.984 & 0.51 & 10.6 \\
        & \textit{SR-CEM} & \textbf{0.431} & \textbf{0.930} & \textbf{0.996} & \textbf{0.37} & \textbf{7.36} \\ \hline
\end{tabular}
\label{tab:ctc_confidence}
\vspace{-2mm}
\end{table}

\vspace{-2mm}
\subsection{Experiment with RNN-T Model}

We also test the methods in the RNN-Transducer (RNN-T) setting. The RNN-T model is trained from scratch on LS train-clean-360.


Tab.~\ref{tab:rnnt_confidence} shows RNN-T results. SR-CEM achieves best token-level performance (NCE 0.353, MCE 4.20\% vs. 21.45\% softmax) and lowest word-level MCE (11.95\% vs. 25.58\% E2EXformer, 25.56\% softmax), confirming effectiveness across attention, CTC, and transducer architectures.

\begin{table}[tb]
\vspace{-2mm}
\centering
\caption{Token and word-level confidence results for RNN-Transducer on the LS test-clean. All CEMs trained and LS train-clean-100. ASR trained on LS train-clean-360.}
\begin{tabular}{l|l|rrrrr}
\hline
Level   & Method       & NCE$\uparrow$ & ROC$\uparrow$ & PR$\uparrow$ & ECE$\downarrow$ & MCE$\downarrow$ \\ \hline
\multirow{4}{*}{Token} 
        & Softmax      & 0.105 & 0.894 & \textbf{0.992} & 3.14 & 21.45 \\
        & MLPCEM       & 0.300 & 0.874 & 0.989 & 0.64 & 20.69 \\
        & Xformer      & 0.289 & 0.863 & 0.988 & 0.56 & 11.56 \\
        & TruCLeS    & -0.023 & 0.635 & 0.965 & 2.62 & 10.76 \\
        & \textit{SR-CEM} & \textbf{0.353} & \textbf{0.903} & \textbf{0.992} & \textbf{0.39} & \textbf{4.20} \\ \hline
\multirow{5}{*}{Word} 
        & Softmax      & 0.179 & \textbf{0.908} & 0.994 & 2.57 & 25.56 \\
        & MLPCEM       & 0.311 & 0.873 & 0.991 & 0.77 & 20.58 \\
        & Xformer      & 0.301 & 0.865 & 0.990 & 0.88 & 14.64 \\
        & E2EXformer   & \textbf{0.337} & \textbf{0.908} & \textbf{0.995} & \textbf{0.57} & 25.58 \\
        & TruCLeS    & -0.026 & 0.625 & 0.964 & 2.56 & 13.37 \\
        & \textit{SR-CEM} & 0.142 & 0.768 & 0.984 & 0.71 & \textbf{11.95} \\ \hline
\end{tabular}
\label{tab:rnnt_confidence}
\vspace{-2mm}
\end{table}

\vspace{-2mm}
\subsection{Experiment on Out-of-Domain Data}



To assess target-domain training, we train SR-CEM on 20k CV utterances while keeping the ASR model unchanged. Tab.~\ref{tab:us_confidence} shows SR-CEM achieves MCE of 6.67\% (token) and 12.15\% (word) compared to 34.77\% and 47.21\% for softmax, confirming benefits from domain-matched training data.

\begin{table}[tb]
\vspace{-2mm}
\centering
\caption{Token and word-level confidence results on the CV test set. All CEMs trained on the CV validation set. ASR trained on CV train set.}
\begin{tabular}{l|l|rrrrr}
\hline
Level   & Method       & NCE$\uparrow$ & ROC$\uparrow$ & PR$\uparrow$ & ECE$\downarrow$ & MCE$\downarrow$ \\ \hline
\multirow{4}{*}{Token} 
        & Softmax      & -0.029        & 0.853             & 0.926             & 15.99          & 34.77         \\
        & MLPCEM       & 0.310        & 0.860             & 0.926             & 2.33          & 7.06         \\
        & Xformer      & 0.297        & 0.856             & 0.925             & 3.96          & 13.21         \\
        & \textit{SR-CEM}       & \textbf{0.339} & \textbf{0.872}   & \textbf{0.937}    & \textbf{2.26} & \textbf{6.67} \\ \hline
\multirow{5}{*}{Word} 
        & Softmax      & -0.485 & 0.792   & 0.842    & 23.84          & 47.21         \\
        & MLPCEM       & 0.073        & 0.780             & 0.825             & 11.65          & 16.11         \\
        & Xformer      & 0.062        & 0.776             & 0.826             & 13.22          & 21.78         \\
        & E2EXformer   & 0.302        & 0.862             & 0.919    & 5.14 & 12.77         \\
        & \textit{SR-CEM}       & \textbf{0.412}        & \textbf{0.891}             & \textbf{0.920}             & \textbf{2.27}          & \textbf{12.15}\\ \hline
\end{tabular}
\label{tab:us_confidence}
\vspace{-2mm}
\end{table}

\vspace{-2mm}
\subsection{Experiment on Dutch Language}




Our method is language-agnostic, as it does not rely on language-specific designs. To validate this property, we evaluated it with a Dutch ASR model~\cite{poncelet2025leveraging} trained on a large media subtitle corpus.

At the token level, SR-CEM achieves the lowest calibration errors (ECE 0.15, MCE 4.41) while maintaining strong ROC and PR scores. At the word level, SR-CEM provides the best calibration with MCE of 6.26\%, compared with 35.85\% for softmax and 7.46\% for Xformer. The results confirm that SR-CEM generalizes well across languages and consistently improves calibration quality.

\begin{table}[tb]
\vspace{-2mm}
\centering
\caption{Token and word-level confidence results for Dutch ASR on the CGN test set. All CEMs trained on CGN validation set. ASR trained on CGN train set\cite{poncelet2025leveraging}.}
\begin{tabular}{l|l|rrrrr}
\hline
Level   & Method       & NCE$\uparrow$ & ROC$\uparrow$ & PR$\uparrow$ & ECE$\downarrow$ & MCE$\downarrow$ \\ \hline
\multirow{4}{*}{Token} 
        & Softmax      & -0.117        & 0.798             & 0.994             & 1.46          & 73.05         \\
        & MLPCEM       & 0.202        & 0.825             & 0.989             & 0.83          & 17.27         \\
        & Xformer      & \textbf{0.259} & 0.768            & 0.960             & 1.59          & 12.79         \\
        & \textit{SR-CEM} & 0.196      & \textbf{0.847}    & \textbf{0.996}    & \textbf{0.15} & \textbf{4.41} \\ \hline
\multirow{5}{*}{Word} 
        & Softmax      & -0.333        & \textbf{0.840}    & \textbf{0.984}    & 5.31          & 35.85         \\
        & MLPCEM       & 0.166         & 0.808             & 0.980             & 2.13          & 33.74         \\
        & Xformer      & 0.040         & 0.669             & 0.979             & \textbf{0.60} & 7.46          \\
        & E2EXformer   & 0.147         & 0.839             & 0.983             & 3.24          & 17.58         \\
        & \textit{SR-CEM} & \textbf{0.230} & 0.825         & 0.980             & 0.95          & \textbf{6.26} \\ \hline
\end{tabular}
\label{tab:dutch_confidence}
\vspace{-2mm}
\end{table}

\vspace{-2mm}
\subsection{Experiment on Noisy Speech}


\begin{table}[tb]
\vspace{-2mm}
\centering
\caption{Token and word-level confidence results on the APT test set. All CEMs trained on the APT validation set. ASR fine-tuned on APT train set.}
\begin{tabular}{l|l|rrrrr}
\hline
Level   & Method       & NCE$\uparrow$ & ROC$\uparrow$ & PR$\uparrow$ & ECE$\downarrow$ & MCE$\downarrow$ \\ \hline
\multirow{4}{*}{Token} 
        & Softmax      & -94.241 & 0.524   & 0.949    & 95.07         & 95.08         \\
        & MLPCEM       & 0.349 & 0.900 & \textbf{0.993} & 0.77 & 12.12         \\
        & Xformer      & 0.326 & 0.890 & 0.992 & 0.43 & 11.85         \\
        & \textit{SR-CEM}       & \textbf{0.364}       & \textbf{0.907}            & \textbf{0.993}            & \textbf{0.32}         & \textbf{5.99} \\ \hline
\multirow{5}{*}{Word} 
        & Softmax      & 0.2484 & \textbf{0.926}    & \textbf{0.996}     & 2.54          & 24.76        \\
        & MLPCEM       & -0,416 & 0.535 & 0.930 & 6.78 & 82.66         \\
        & Xformer      & -0.395 & 0.533 & 0.931 & 7.00 & 81.33        \\
        & E2EXformer   & \textbf{0.394} & 0.924 & \textbf{0.996} & 0.65 & 9.50       \\
        & \textit{SR-CEM}       & 0.346       & 0.891            & 0.992            & \textbf{0.64}         & \textbf{5.58} \\ \hline
\end{tabular}
\label{tab:apt_confidence}
\vspace{-2mm}
\end{table}


To evaluate our method in practical noisy environments, we fine-tuned the backbone ASR on Libri-Adapt, achieving 6.3\% WER and 7.4\% TER.

The results in Tab.~\ref{tab:apt_confidence} show that confidence estimation on noisy data is more challenging than on clean speech. At the token level, SR-CEM achieves the best performance across all metrics (MCE 5.99\%). At the word level, while E2EXformer shows strong discrimination, SR-CEM provides the most consistent calibration with the lowest ECE (0.64\%) and MCE (5.58\%). These findings highlight SR-CEM's robustness in practical noisy scenarios.

\vspace{-2mm}
\subsection{Experiment on Conversational Speech}
To further evaluate our method in a more real-word setting, we fine tune the backbone CTC/Attention ASR model on the CHiME-6 train set. 

Tab.~\ref{tab:chime6} shows SR-CEM's performance on conversational speech. SR-CEM achieves the best calibration at both token level (MCE 7.51\% vs. 24.69\% for softmax) and word level (MCE 7.46\%, ECE 3.42\%), demonstrating robustness on realistic spontaneous speech alongside the read and noisy speech results.

\begin{table}[tb]
\vspace{-2mm}
\centering
\caption{Token and word-level confidence results on the CHiME-6 test set. All CEMs trained on the CHiME-6 validation set. ASR fine-tuned on CHiME-6 train set.}
\begin{tabular}{l|l|rrrrr}
\hline
Level   & Method       & NCE$\uparrow$ & ROC$\uparrow$ & PR$\uparrow$ & ECE$\downarrow$ & MCE$\downarrow$ \\ \hline
\multirow{4}{*}{Token} 
        & Softmax      & 0.015 & 0.784   & 0.838    & 13.63         & 24.69         \\
        & MLPCEM       & 0.173 & 0.793 & 0.846 & 3.49 & 22.65         \\
        & Xformer      & 0.183 & 0.789 & 0.841 & 3.38 & 12.36         \\
        & \textit{SR-CEM}       & \textbf{0.207}       & \textbf{0.798}            & \textbf{0.850}            & \textbf{1.93}         & \textbf{7.51} \\ \hline
\multirow{5}{*}{Word} 
        & Softmax      & 0.189 & 0.828    & 0.892     & 10.79          & 23.59        \\
        & MLPCEM       & 0.216 & 0.817 & 0.871 & 3.69 & 19.87         \\
        & Xformer      & 0.219 & 0.810 & 0.867 & 3.88 & 10.00        \\
        & E2EXformer   & 0.218 & 0.814 & 0.869 & 3.98 & 14.12       \\
        & \textit{SR-CEM}       & \textbf{0.252}       & \textbf{0.829}            & \textbf{0.885}            & \textbf{3.42}         & \textbf{7.46} \\ \hline
\end{tabular}
\label{tab:chime6}
\vspace{-2mm}
\end{table}

\subsection{Discussion}

Our evaluation reveals several consistent patterns. SR-CEM prioritizes calibration over maximal discrimination, achieving competitive ROC/PR while dramatically reducing MCE—a trade-off valuable for threshold-based applications. This pattern holds across all architectures (hybrid, attention-only, CTC-only, RNN-T), with CTC-only showing the best calibration (MCE 3.27\%), likely because frame-independent assumptions reduce the local-global mismatch in autoregressive decoding.

Domain shift affects all methods, but SR-CEM maintains its relative advantage on out-of-domain CV (MCE 19.59\% vs. 34.77\% softmax), improving substantially when trained on the target domain (6.67\%). Language-agnostic behavior is confirmed by Dutch results (MCE 4.41\% token, 6.26\% word) mirroring English performance. Acoustic robustness extends to noisy (5.99\%/5.58\%) and conversational (7.51\%/7.46\%) conditions, with conversational speech presenting the greatest challenge due to spontaneous disfluencies and background noise.

Architecture-dependent baselines show inconsistent performance. MLPCEM improves ECE but not MCE, while TruCLeS shows weak discrimination despite being applicable to multiple architectures. Our results highlight MCE's importance: while many methods achieve low average error, SR-CEM's consistent 50-70\% MCE reduction addresses worst-case bins critical for practical deployment.

The word-level performance gap (8.17\% vs. 4.50\% token-level MCE) reflects the absence of top-K features. Future work could incorporate N-best list features or beam statistics, though this would come at a computational cost.

\section{CONCLUSION AND FURTHER RESEARCH}
We proposed SR-CEM, a lightweight confidence estimation module that leverages score and rank information from beam search decoding. Unlike previous methods, SR-CEM is independent of the ASR model architecture and relies solely on information available during decoding. Our experiments show that SR-CEM performs strongly on both in-domain and out-of-domain datasets, and consistently outperforms multiple baseline methods across diverse evaluation metrics. In particular, it is effective in reducing Maximum Calibration Error (MCE), which is critical for reliable downstream use of ASR outputs. To further validate its robustness, we evaluated SR-CEM against multiple baselines across different ASR architectures, datasets, languages, and noisy and conversational conditions. These results demonstrate both the effectiveness and the generality of the method.

SR-CEM has several limitations. It requires beam search decoding, as greedy decoding or aggressive pruning reduces feature informativeness. It fails when there is serious mismatch between training and test conditions, such as training on clean speech but testing on noisy conditions, or encountering out-of-domain data with new vocabulary and acoustic patterns (CV: MCE 19.59\% vs. 4.50\% in-domain). These mismatches cause beam search score distribution shifts that degrade calibration. SR-CEM relies on the single-best hypothesis, ignoring beam alternatives. When multiple hypotheses have similar scores, it cannot detect this uncertainty. Word-level calibration remains weaker than token-level (8.17\% vs. 4.50\% MCE) due to absent top-K features. While SR-CEM's architecture is language-agnostic, learned calibration curves are not. Cross-language deployment requires retraining.

Confidence estimation in ASR remains fragmented and lacks a unified framework. This work moves toward an architecture-agnostic approach, though extending confidence estimation across models is still an open direction. Future work could address these limitations through training on diverse speech data to learn more robust confidence patterns, incorporating N-best features or beam diversity measures, developing domain-adaptive calibration methods, and exploring context-aware word-level aggregation strategies. Although our experiments used a hybrid Attention/CTC model, the idea behind SR-CEM generalizes to setups that integrate multiple knowledge sources, such as lexical constraints or large language model scores.

\bibliographystyle{IEEEtran}
\bibliography{references}
\vspace{-3mm}

\begin{IEEEbiography}[{\includegraphics[width=1in,height=1.25in,clip,keepaspectratio]{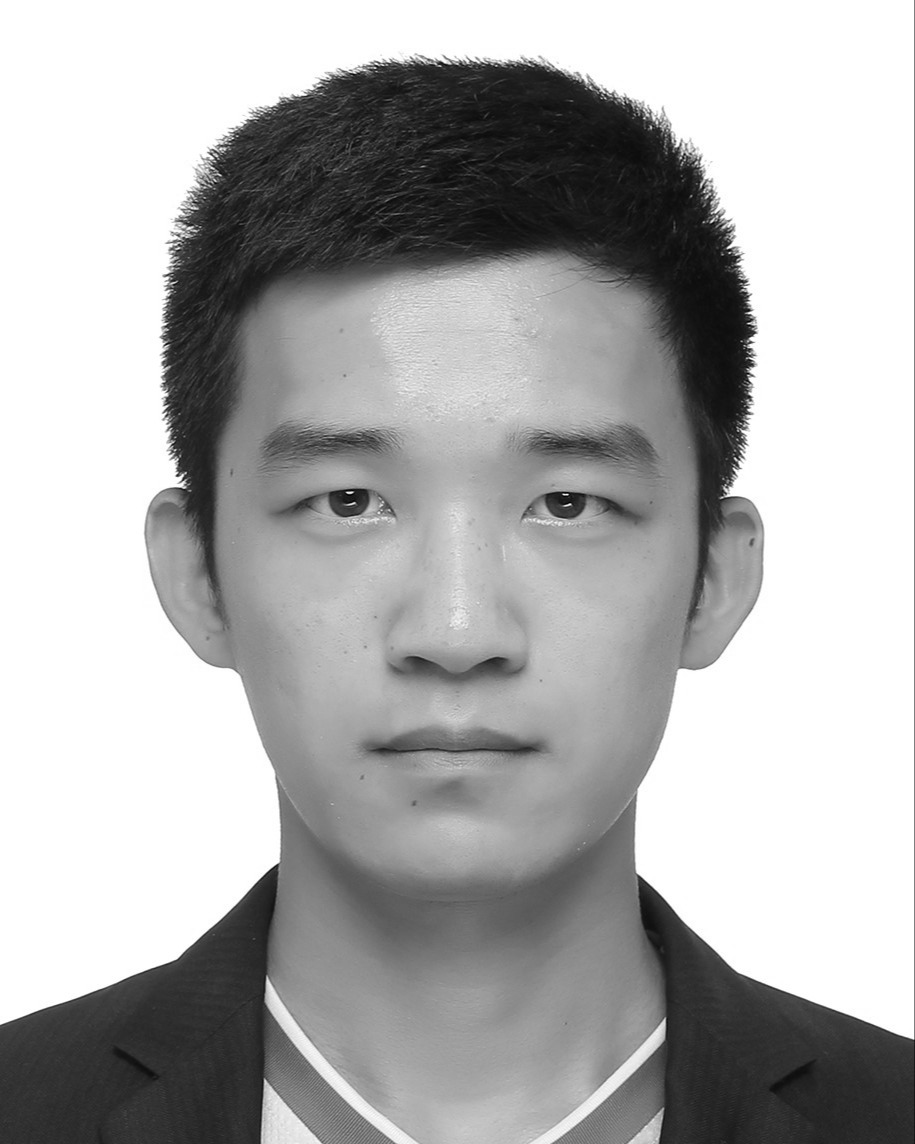}}]
{Yichen Jia}~(Member, IEEE)~received the B.S. degree in financial engineering from Sichuan University, Chengdu, China, in 2021, and the M.S. degree in statistics and data science from KU Leuven, Leuven, Belgium, in 2024.

He is currently a Ph.D. student at the (Processing Speech and Images) PSI Lab, Department of Electrical Engineering (ESAT), KU Leuven, Leuven, Belgium. His current research interests include automatic speech recognition, confidence estimation, streaming ASR, and LLM-ASR.
\end{IEEEbiography}

\vspace{-3mm}

\begin{IEEEbiography}[{\includegraphics[width=1in,height=1.25in,clip,keepaspectratio]{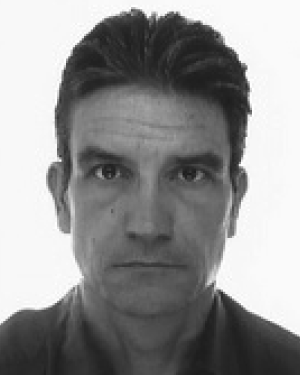}}]
{Hugo Van hamme}~(Senior Member, IEEE) received the master’s degree in engineering (burgerlijk ingenieur) from Vrije Universiteit Brussel (VUB), Brussels, Belgium, in 1987, the M.Sc.\ degree from Imperial College London, London, U.K., in 1988, and the Ph.D.\ degree in electrical engineering from VUB, in 1992. 

From 1993 to 2002, he was with LH Speech Products and ScanSoft, initially as a Senior Researcher and later as a Research Manager. Since 2002, he has been a Professor with the Department of Electrical Engineering ESAT-PSI, KU Leuven, Leuven, Belgium. His main research interests include automatic speech assessment, assistive speech technology, source separation, noise-robust speech recognition, models of language acquisition, and computer-assisted learning.
\end{IEEEbiography}

\end{document}